Growth and characterization of (La,Ce)OBiS$_2$ single crystals

Yuji Hanada[1], Masanori Nagao[1*], Akira Miura[2], Yuki Maruyama[1], Satoshi Watauchi[1], Yoshihiko Takano[3,4], Isao Tanaka[1]

[1]*University of Yamanashi, 7-32 Miyamae, Kofu, Yamanashi 400-8511, Japan*

[2]*Hokkaido University, Kita-13 Nishi-8, Kita-ku, Sapporo, Hokkaido 060-8628, Japan*

[3]*University of Tsukuba, 1-1-1 Tennodai, Tsukuba, Ibaraki 305-8577, Japan*

[4]*MANA, National Institute for Materials Science, 1-2-1 Sengen, Tsukuba, Ibaraki 305-0047, Japan*

[*]Corresponding Author

Masanori Nagao

Postal address: University of Yamanashi, Center for Crystal Science and Technology Miyamae 7-32, Kofu, Yamanashi 400-8511, Japan

Telephone number: (+81)55-220-8610

Fax number: (+81)55-220-8270

E-mail address: mnagao@yamanashi.ac.jp

**Abstract**

(La,Ce)OBiS$_2$ single crystals are successfully grown using CsCl flux. The obtained single crystals have a plate-like shape with a size of 0.5–2.0 mm and a well-developed *ab*-plane. The thickness of the crystals increases with a decrease in the Ce content. The (La,Ce)OBiS$_2$ single crystals exhibit superconducting properties. Moreover, the superconducting transition temperature increases with an increase in the Ce content. The obtained (La,Ce)OBiS$_2$ single crystals exhibit a tetragonal structure and the valence state of Ce in the crystals is determined to be mixed with Ce$^{3+}$ and Ce$^{4+}$ based on X-ray absorption fine structure spectroscopy.

**Main text**

**Introduction**

Cu- and Fe-based superconductors have a high superconducting transition temperature. These materials both have a layered structure consisting of a charge reservoir layer and a superconducting layer. As such, there is interest in superconductors with layered structures. $BiS_2$-based superconductors also have a layered structure consisting of a charge reservoir layer and a $BiS_2$ superconducting layer. A variety of $BiS_2$-based superconductors has been reported including $Bi_4O_4S_3$,[1] $R(O,F)BiS_2$ ($R$=La, Ce, Pr, Nd, Yb),[2–6] $A$F$BiS_2$ ($A$=Sr, Eu),[7–9] (La,$M$)$BiS_2$ ($M$=Ti, Zr, Hf, Th),[10] $Eu_3F_4Bi_2S_4$,[11] $Bi(O,F)BiS_2$.[12,13] $R(O,F)BiS_2$ has a layered structure that consists of a charge reservoir $R(O,F)$ and $BiS_2$ layers. This material behaves as a superconductor when the electron carrier is supplied to the $BiS_2$ layer. Typical doping is facilitated by F substitution of O sites in $ROBiS_2$. Thus, $BiS_2$-based superconductors are similar to $R(O,F)FeAs$.[14,15]

Ce compounds exhibit very important properties. For examples, $CeCu_2Si_2$ have both antiferromagnetic and superconducting properties [16] while $Ce(O,F)BiS_2$ was reported to exhibit both ferromagnetism and superconductivity.[17] $CeOBiS_2$ exhibits superconductivity without F-substitution. $Ce^{3+}$ and $Ce^{4+}$ both exist in $CeOBiS_2$, which generates electron carriers.[18,19] In contrast, $PrOBiS_2$, which has only $Pr^{3+}$ ions, does not exhibit a superconducting transition.[18,20] In terms of the crystal system of $BiS_2$-based compounds, superconducting $CeOBiS_2$ and $R(O,F)BiS_2$ are tetragonal structures.[19] However, non-superconducting $LaOBiS_2$ and $PrOBiS_2$ are monoclinic structures.[20,21] Further studies of tetragonal and monoclinic structures on superconducting properties have been reported for $(Ce,Pr)OBiS_2$.[20] However, the effect of f electrons on Ce and Pr is not clear. The simpler $(La,Ce)OBiS_2$ system is expected to clarify the effect of f electrons on the superconducting and magnetic properties since only $Ce^{3+}$ has f electrons in $La^{3+}$, $Ce^{3+}$, and $Ce^{4+}$.

In this investigation, we grew $(La,Ce)OBiS_2$ single crystals and examined their crystal system, Ce valence state, and superconducting properties to evaluate the effect of Ce substitution.

**Methods**

Single crystals of $(La,Ce)OBiS_2$ were grown using CsCl flux.[18,22-24] The raw materials of $La_2S_3$, $Ce_2S_3$, $Bi_2O_3$, $Bi_2S_3$ were weighed to have a nominal composition of $La_{1-x}Ce_xOBiS_2$ ($0.1 \leq x \leq 1.0$). The mixture of the raw materials (0.8 g) and CsCl flux (5.0 g) was ground using a mortar and then sealed in an evacuated quartz tube (~10 Pa). The quartz tube was heated at 1000 °C except for $x = 1.0$ (950 °C) for 10 h, followed by cooling to 650 °C at a rate of 1 °C/h, then the sample was cooled down to room temperature in the furnace. The heated quartz tube was opened to air and the obtained materials were washed and filtered to remove the CsCl flux using distilled water.

The compositional ratio of the single crystals was evaluated using energy dispersive X-ray spectrometry (EDS) (Bruker, Quantax 70) associated with the observation of the microstructure, based on scanning electron microscopy (SEM) (Hitachi High-Technologies, TM3030). The obtained compositional values were normalized using $C_S = 2$ (S composition was 2), with La, Ce and Bi compositions ($C_{La}$, $C_{Ce}$, and $C_{Bi}$) measured to a precision of two decimal places. For the La and Ce compositions, the $C_{La}$ to $C_{Ce}$ ratio was normalized using $C_{La} + C_{Ce} = 1$ to clarify the relationship between the nominal and analytical compositions. Identification of the crystal structure and the $c$-axis lattice parameters of the grown crystals was performed using X-ray diffraction (XRD) (Rigaku MultiFlex) with CuK$\alpha$ radiation.

Structural analysis of (La,Ce)OBiS$_2$ ($x$ = 0.25, 0.50, 1.00) single crystals was conducted using synchrotron powder X-ray diffraction measurements on crushed single crystals. Synchrotron powder X-ray diffraction measurements were performed at 150 K in SPring-8 using the BL02B2 beamline with the approval of 2018A0074. Rietveld refinements were performed using RIETAN-FP.[25]

The Resistivity-temperature ($\rho$-$T$) characteristics of the obtained single crystals were determined using the standard four-probe method with a constant current density ($J$) mode using a physical property measurement system (Quantum Design; PPMS DynaCool). The electrical terminals were fabricated using Ag paste. $\rho$–$T$ characteristics in the temperature range of 0.20–15 K were determined based using the adiabatic demagnetization refrigerator (ADR) option for the PPMS. A magnetic field of 3 T at 1.9 K was utilized to operate the ADR, which was subsequently removed. Consequently, the temperature of the sample decreased to approximately 0.20 K. The measurement of $\rho$–$T$ characteristics was initiated at the lowest temperature (~0.20 K), which was spontaneously increased to 15 K. The superconducting transition temperature ($T_c$) was estimated from the $\rho$–$T$ characteristics. The transition temperature corresponding to the onset of superconductivity ($T_c^{onset}$) is defined as the temperature at which deviation from linear behavior is observed in the normal conducting state for the $\rho$–$T$ characteristics. The zero resistivity ($T_c^{zero}$) is determined as the temperature at which resistivity is below 300 $\mu\Omega$ cm.

The valence state of the Cerium component in the obtained single crystals was estimated using X-ray absorption fine structure (XAFS) spectroscopy analysis of Ce-$L_3$ edges using an Aichi XAS beamline with synchrotron X-ray radiation (BL11S2: Experimental No.201801025). To prepare samples for the XAFS spectroscopy measurements, the obtained single crystals were ground and mixed with boron nitride (BN) powder, before being pressed into a pellet of 4 mm diameter.

**Results and Discussion**

Figure 1(a) and (b) show typical SEM images of (La,Ce)OBiS$_2$ single crystals. The obtained single crystals had a plate-like shape with a size of 0.5–2.0 mm and 150–800 μm in thickness. The thickness of the single crystals tended to decrease with increasing Ce content. Table I shows La, Ce and Bi compositions of the grown (La,Ce)OBiS$_2$ single crystals that were analyzed via EDS. It was confirmed that the composition was similar to the nominal composition for all the samples.

Figure 2 shows the XRD patterns of a well-developed plane in single crystals of La$_{1-x}$Ce$_x$OBiS$_2$ ($x$=0.1–1.0). The XRD patterns indicate that the *ab*-plane is well-developed for single crystals because only 00*l* diffraction peaks are found, which is similar to other BiS$_2$-based materials.[18,24] The results of EDS and XRD confirm that (La,Ce)OBiS$_2$ single crystals were successfully grown. The *c*-axis lattice parameter was calculated from the 00$\underline{11}$ peak and is shown in the inset of Figure 2. The *c*-axis lattice parameter decreases with an increase of the Ce content. These results are attributed to the smaller ionic radius of Ce ($Ce^{3+}$:1.034 Å, $Ce^{4+}$:0.80 Å) compared to that of La ($La^{3+}$:1.061 Å).[26]

Figure 3 shows synchrotron X-ray diffraction patterns at 150 K for the crushed single crystals for *x* = 0.25, 0.50 and 1.00. All samples were indexed as having a tetragonal structure. The inset of Figure 3 shows enlarged patterns of the 200 and 201 diffraction peaks. The width of these peaks is comparable, therefore, these are tetragonal structures; 200 diffraction peaks of monoclinic $PrOBiS_2$ are clearly separated at 100 K and become broad at 300 K.[20]

Figure 4 shows the resistivity-temperature (*ρ-T*) characteristics for the obtained single crystals. The single crystals for *x* = 0.25 do not exhibit superconductivity down to 0.2 K, although other single crystals show superconducting transition. The superconducting transition temperature increases with the increase of the Ce content.

Figure 5 shows Ce-$L_3$ edge XAFS spectra at room temperature for the single crystals of $Ce_2S_3$($Ce^{3+}$) and $CeO_2$($Ce^{4+}$). An increase in the Ce content results in a reduction of the peak at 5740 eV, which is characteristic of $Ce^{4+}$. The Ce valence state in (La,Ce)$OBiS_2$ single crystals was a mixture of $Ce^{3+}$ and $Ce^{4+}$. This result consists in the angle-resolved photoemission spectroscopy (ARPES) study of $CeOBiS_2$ ($x = 1.0$) single crystals.[27] Therefore, these valence states are not only attributed to the surface state but also to the bulk property.

The correlation between superconductivity and crystal structure (tetragonal or monoclinic), carrier concentration, valence fluctuation, effect of f electrons, and chemical pressure have been discussed. In previous reports on crystal structure, $LaOBiS_2$ exhibited a monoclinic structure with non-superconductivity[21], while $CeOBiS_2$ showed a tetragonal structure with superconductivity.[19] However, in this report on $(La,Ce)OBiS_2$, superconductivity disappeared down to 0.25 K at $x = 0.25$, even though this crystal has a tetragonal structure. Therefore, the appearance of superconductivity is not strongly correlated with crystal structure.

Valence fluctuation is also an important factor for superconductivity in $BiS_2$-based compounds. For example, $CeOBiS_2$ [18], $EuFBiS_2$ [9] and $Eu_3F_4Bi_2S_4$ [11] exhibited superconductivity due to the mixed valence state of Ce or Eu. $PrOBiS_2$ exhibited little valence fluctuation which was almost $Pr^{3+}$ including little $Pr^{4+}$.[28] Thus $PrOBiS_2$ is non-superconductor. Additionally, F-doped $PrOBiS_2$ becomes superconductor [29] with only $Pr^{3+}$ state,[30] whose superconductivity is induced by the F-doping. The valence state of rare earth elements exhibits a wide variety of behavior in $BiS_2$-based compounds. In comparison, in the investigation of the correlation between valence fluctuation and superconductivity in (La,Ce)OBiS$_2$ single crystals, the mixed valence states of Ce were determined using XAFS spectroscopy (Figure 5). To discuss the carrier concentration in further details, the valence fluctuation and chemical pressure are considered. Figure 6(a) shows the ion concentration ratio of La$^{3+}$, Ce$^{3+}$, and Ce$^{4+}$ of (La,Ce)OBiS$_2$ single crystals derived from XAFS spectroscopy. It is evident that while the ion concentration ratio of Ce$^{4+}$ is almost constant with for different Ce content, Ce$^{3+}$ increases and La$^{3+}$ decreases with increasing Ce content. Assuming that the carrier concentration is only determined by the valence of the rare earth elements, the carrier concentration is almost constant. However, this assumption requires further investigation because it has not been proven that the carrier concentration of the $R$O and BiS$_2$ layers is the same. Further investigation is warranted in several characterization; for example, in hole measurement.

Figure 6(b) shows the average ionic radii that were calculated based on the ratio and ionic radii of La$^{3+}$, Ce$^{3+}$, and Ce$^{4+}$.[26] The averaged ionic radius decreases with increasing Ce contents Therefore, this implies that an increase in the superconducting transition temperature can be considered in terms of the chemical pressure induced by Ce$^{3+}$, given that a constant Ce$^{4+}$ concentration ensures similar carrier concentration in all samples. Given that this trend is same as that observed in other systems including (Ce,Pr)OBiS$_2$ and (Ce,Nd)O$_{0.5}$F$_{0.5}$BiS$_2$,[20,31] the f electrons of Ce$^{3+}$ is not a dominant factor with respect to the increase of superconductivity. It is indisputable that the number of carriers supplied to the BiS$_2$ layer increases due to the chemical pressure.

Superconductivity due to bulk or filamentary characteristics is still an open question. For instance, CeOBiS$_2$ ($x = 1.0$) single crystals in an ARPES study sample [27] was determined to be non-bulk-superconductor based on specific heat measurements.[32] Further investigations of the superconducting volume fraction of (La,Ce)OBiS$_2$ single crystals are required. For example, magnetization measurements at a lower temperature will be a powerful way of identifying bulk or filamentary superconductors.

**Conclusions**

La$_{1-x}$Ce$_x$OBiS$_2$ ($0.1 \leq x \leq 1.0$) single crystals were successfully grown using CsCl flux. Superconductivity was indicated by $x \geq 0.5$ and the superconducting transition temperature tends to increase with an increase of the Ce content. Superconductivity was not demonstrated at $x = 0.25$ although the crystal system was tetragonal. This result confirmed that superconductivity is not related to the crystal system. In the $R$OBiS$_2$ group, the non-superconducting materials of the tetragonal structure were not reported in previous studies. These results assumed that an increase of the superconducting transition temperature is not due to carrier doping of $Ce^{4+}$, which is attributed to the chemical pressure effect with increasing $Ce^{3+}$.

**Acknowledgments**

Synchrotron powder X-ray diffraction measurements were performed at the BL02B2 of SPring-8 with the approval of 2018A0074. The XAFS spectroscopy experiments were conducted at the BL11S2 of Aichi Synchrotron Radiation Center, Aichi Science & Technology Foundation, Aichi, Japan (Experimental No. 201801025). This work was partial supported by JSPS KAKENHI (Grant-in-Aid for Scientific Research (C)) Grant Number 19K05248.

We would like to thank Editage (www.editage.jp) for English language editing.

Table I. Nominal Ce composition ($x$), the analytical La, Ce and Bi compositions ($C_{La}$, $C_{Ce}$, and $C_{Bi}$) in the obtained single crystals. The analytical $C_{La}$ and $C_{Ce}$ compositions were normalized by total $C_{La} + C_{Ce} = 1$ and $C_S = 2$.

| | | | Nominal Ce composition ($x$) in La$_{1-x}$Ce$_x$OBiS$_2$ | | | | | | |
|---|---|---|---|---|---|---|---|---|---|
| | | | 0.10 | 0.25 | 0.50 | 0.75 | 0.80 | 0.90 | 1.00 |
| Analytical composition (normalized by $C_S$=2) | $C_{La} + C_{Ce} = 1$ | $C_{La}$ | 0.912 ±0.006 | 0.754 ±0.004 | 0.517 ±0.007 | 0.249 ±0.003 | 0.224 ±0.007 | 0.111 ±0.016 | 0 |
| | | $C_{Ce}$ | 0.089 ±0.006 | 0.246 ±0.004 | 0.483 ±0.007 | 0.751 ±0.003 | 0.776 ±0.007 | 0.889 ±0.016 | 1.000 ±0.008 |
| | | $C_{Bi}$ | 0.999 ±0.012 | 0.994 ±0.016 | 0.986 ±0.008 | 0.988 ±0.017 | 0.997 ±0.009 | 0.965 ±0.013 | 0.995 ±0.011 |

**Figure captions**

Figure 1. Typical SEM images of $La_{1-x}Ce_xOBiS_2$ single crystals grown from (a) $x=0.25$ and (b) $x=0.90$.

Figure 2. XRD patterns of a well-developed plane of the single crystals grown from the nominal composition of $La_{1-x}Ce_xOBiS_2$ ($x = 0.10–1.0$). The inset represents the nominal Ce composition ($x$) dependence of the $c$-axis lattice parameters.

Figure 3. Synchrotron XRD patterns at 150 K of the grounded single crystals from the nominal composition of $La_{1-x}Ce_xOBiS_2$ ($x = 0.25, 0.50$ and $1.00$). The inset represents a magnification at approximately 200 and 201 peaks.

Figure 4. Resistivity-temperature ($\rho$-$T$) characteristics along the $ab$-plane of the obtained single crystals grown from the nominal composition of $La_{1-x}Ce_xOBiS_2$ ($x = 0.25–1.0$). The inset represents the nominal Ce content ($x$) dependence of the superconducting transition temperature ($T_c^{onset}$ and $T_c^{zero}$). The data for $x = 1.0$ is from Ref. 18.

Figure 5. Ce L3-edge, XAFS spectra obtained at room temperature for the obtained single crystals grown from the nominal composition of $La_{1-x}Ce_xOBiS_2$ ($x = 0.25–1.0$), $Ce_2S_3$, and $CeO_2$.

Figure 6. Nominal Ce contents ($x$) dependence of (a) ion concentration ratio of $La^{3+}$,

$Ce^{3+}$ and $Ce^{4+}$, (b) averaged (La,Ce)-site ionic radii for the obtained single crystals. The ratio of $Ce^{3+}$ and $Ce^{4+}$ in the samples were calculated using linear combination fitting of the reference samples of $Ce_2S_3$($Ce^{3+}$) and $CeO_2$($Ce^{4+}$).

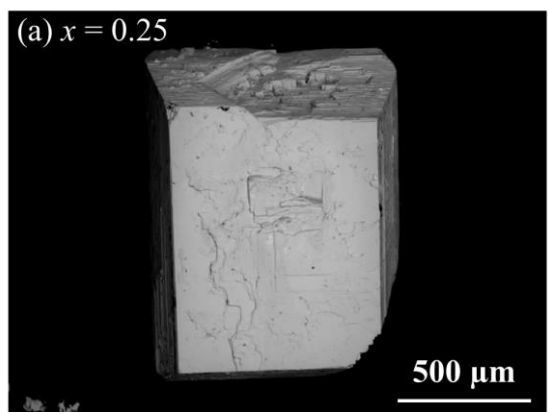 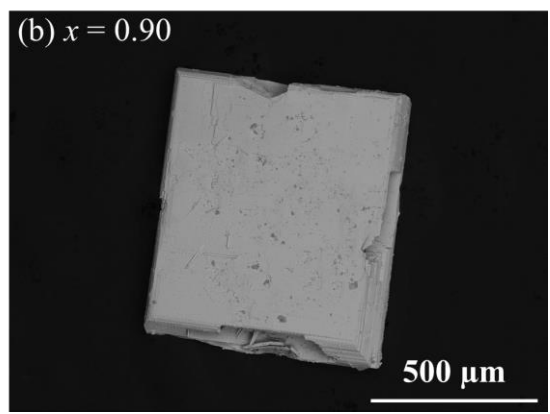

**Figure 1**

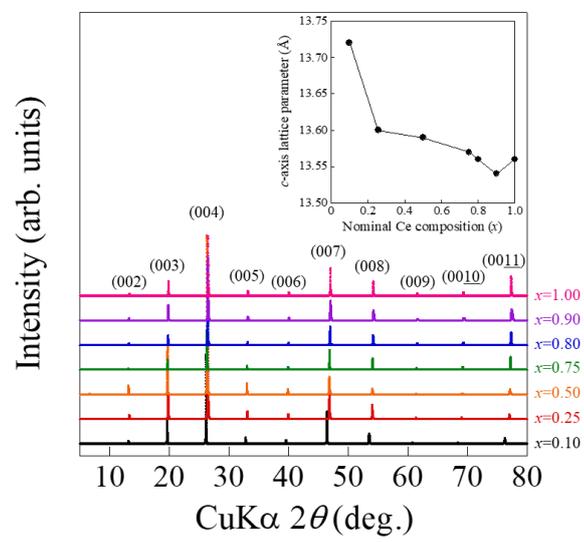

**Figure 2**

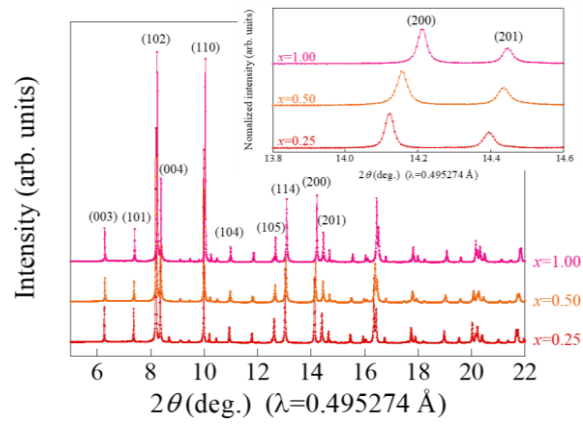

**Figure 3**

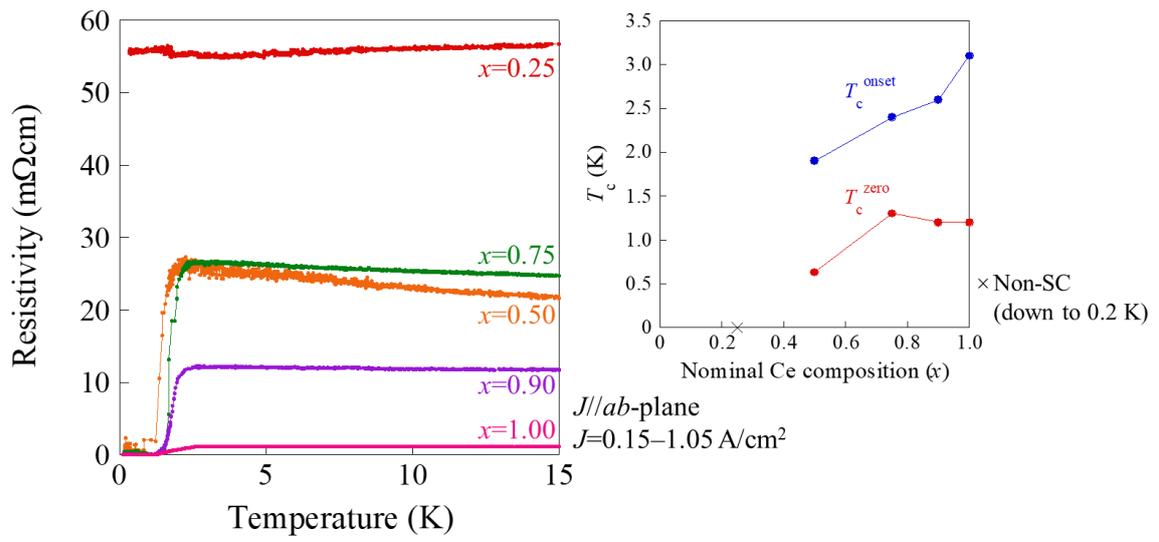

**Figure 4**

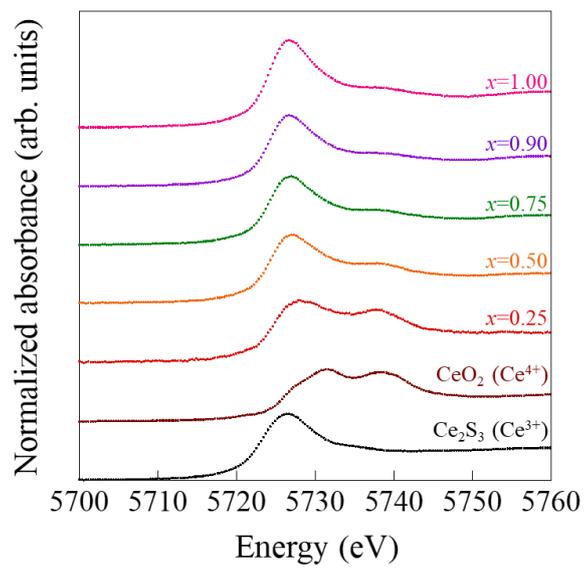

**Figure 5**

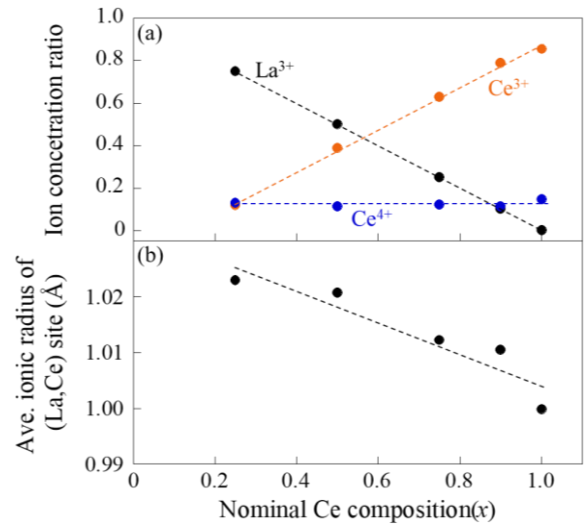

**Figure 6**